# Quantum Critiality and Collective Effects in Low Dimensional Magnet CuGeO$_3$ Probed by High Frequency EPR

S.V Demishev, [1] A.V.Semeno, [1] A.A.Pronin, [1] N.E.Sluchanko, [1] N.A.Samarin, [1] H.Ohta, [2] S.Okubo, [2] M.Kimata, [3] K.Koyama, [4] M. Motokawa, [4] A.V.Kuznetsov [5].

For the CuGeO$_3$ single crystals doped with 1% of Fe the quantum critical behavior in a wide temperature range 1-40 K is reported. The critical exponents for susceptibility along different crystallographic axes are determined: α=0.34 (***B***∥***a*** and ***B***∥***c***) and α=0.31 (***B***∥***b***). The description of the temperature dependences of the line width and g-factor could be obtained in the OA theory assuming that staggered component of the magnetic moment is located predominantly along ***b*** axis. Possible arguments favoring the competition between effects of the staggered field and antiferromagnetic ordering are provided.

**KEY WORDS:** CuGeO$_3$, Quantum critical phenomena, EPR

## 1. INTRODUCTION

Studying of the quantum critical (QC) phenomena becomes a rapidly developing field in condensed matter physics. The QC behavior may appear at special points of various phase diagrams or may be induced by disorder. From experimental point of view the first case corresponds to heavy fermion materials whereas the second is addressed mainly to disordered magnets. In the latter case the ground state of the system for $T<T_G$ is represented by Griffiths phase (GP) [1]. The characteristic feature of the GP is a divergent magnetic susceptibility, which acquires essentially non Curie-Weiss form,

$$\chi(T) \sim 1/T^\alpha, \qquad (1)$$

where α<1 [2]. Recently we have shown that doping with 1% of Fe impurity (S=2) of CuGeO$_3$ induces in high quality single crystals a strong disorder in magnetic subsystem of Cu$^{2+}$ (S=1/2) quantum spin chains and leads to a complete damping of both spin-Peierls and Neel transitions [4-7]. This result has been experimentally confirmed by magnetic susceptibility and specific heat data [4] as well as by studying of the magnetic phase diagram [3]. The QC phenomena including divergent susceptibility with the exponent α=0.36 (Eq. (1)) have been reported below 30-70 K and could be observed at least down to 1.8 K [3-7]. However the question about temperature asymptotic of spin susceptibility at very low temperatures, which is essential for the quantum critical problem, remains open. The second unconsidered up to now aspect is that the critical behavior given by equation (1) have been checked only for the case when external magnetic field ***B*** was parallel to ***a*** crystallographic axis [3-7] and the information for ***B***∥***b*** and ***B***∥***c*** is missing. At the same time the theoretical consideration [1-3] strongly suggests a universal critical behavior in the QC regime [1,2], and hence studying of the various orientation of magnetic field in highly anisotropic material like CuGeO$_3$ may be important for checking of the QC model. In the present work we are aimed on getting experimental answers on the two aforementioned problems.

[1] Low Temperatures and Cryogenic Engineering Dept., General Physics Institute of RAS, Vavilov str. 38, 119991 Moscow, Russia
[2] Molecular Photoscience Research Center, Kobe University, 1-1 Rokkodai, Nada, Kobe 657-8501, Japan
[3] The Graduate School of Science and Technology, Kobe University, 1-1 Rokkodai, Nada, Kobe 657-8501, Japan
[4] Institute for Materials Research, Tohoku University, Sendai 980-8577, Japan
[5] Moscow Engineering Physics Institute, Kashirskoe Shosse, 31, 115409 Moscow, Russia

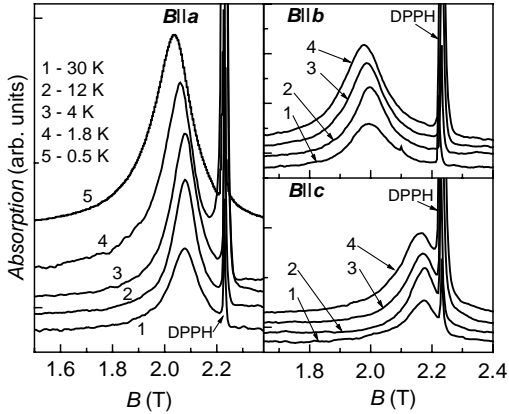

**Fig. 1**. EPR spectra of CuGeO$_3$ doped with 1% of Fe.

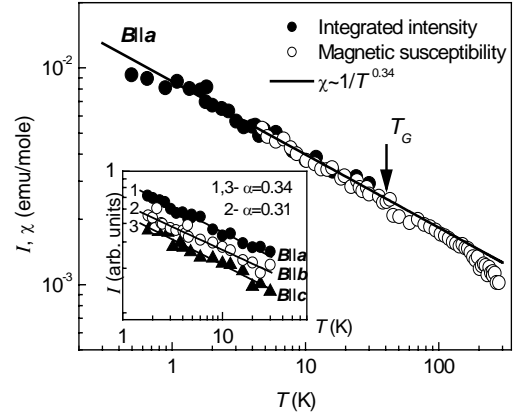

**Fig. 2**. Temperature dependences of the integrated intensity and magnetic susceptibility for the various orientations of magnetic field.

## 2. SAMPLES AND EXPERIMENTAL RESULTS

Spin part of the magnetic susceptibility have been probed by 60 GHz cavity EPR spectrometers. We were able to extend EPR measurements down to 0.5 K using He$^3$ cryostat. For the absolute calibration of $\chi(T)$ the independent measurements of the same crystal have been performed in SQUID magnetometer. The studied single crystals of CuGeO$_3$ contained 1% of Fe impurity. The details about samples preparation and quality control may be found elsewhere [3,8].

The experimental EPR spectra for CuGeO$_3$:Fe (fig. 1) consist of a single line corresponding to resonance on chains of Cu$^{2+}$ ions in agreement with the published data [3-7]. The quality of the spectra has allowed to find a full set of spectroscopic data (line widths $W$, $g$-factors, integrated intensities $I$) for the whole temperature-range studied. The line shape was found to be lorentzian within the experimental accuracy.

Fig. 2 summarize the integrated intensity data; for $B \parallel a$ the EPR $I(T)$ data are placed together with $\chi(T)$ data obtained with the help of the SQUID magnetometer. At measuring frequency 60 GHz for the CuGeO$_3$:Fe the resonant field is relatively high (about 2 T, fig. 1), nevertheless the condition of the linear magnetic response is fulfilled [5] and relation $I(T) \sim \chi(T)$ is valid (fig. 2). We found that the power law (1) describes experimental data in a wide temperature range $1 < T < 40$ K (fig. 2). The corresponding value of the critical exponent is 0.34±0.02 and practically coincide with the data of Ref. 3-6. Below 1 K the $I(T)$ curve tends to saturate, but no sign of antiferromagnetic (AF) transition was found down to the lowest temperature studied. The beginning of the power law provides an estimate of the temperature for the transition into Griffiths phase $T_G \sim 40$ K.

The effects of anisotropy of the $I(T)$ are illustrated by inset in fig. 2. For $B \parallel a$ and $B \parallel c$ the index $\alpha$ is identical: $\alpha = 0.34 \pm 0.02$. In the case $B \parallel b$ the magnitude of $\alpha$ equals $\alpha = 0.31 \pm 0.02$ and somewhat smaller. Nevertheless all observed values are very close and thus strongly support idea about universal behavior of susceptibility in the QC regime. Moreover this result agrees well with the specific heat data reported in [4,6], where the magnetic contribution having form $C_m \sim T^{1-\alpha}$ with $\alpha = 0.35 \pm 0.03$ was found. Indeed the disorder driven QC behavior is characterized be aforementioned power laws for specific heat and susceptibility with the same exponent α [1,2,4,6].

The temperature dependences of the line width and g-factor demonstrate anomalous low temperature growth (fig. 3 and inset in fig. 3), which is qualitatively the same for various orientations of magnetic field with respect to crystal axes. It is worth to note that in pure $CuGeO_3$ line width always decreases with lowering temperature [10,11] and the temperature dependence of the g-factor is different for **B**||**a** **B**||**b** and **B**||**c** (see solid lines in inset at fig. 3).

## 3. DISCUSSION

The observed $W(T)$ and $g(T)$ data can be explained in the Oshikawa-Affleck (OA) theory, which treats EPR in S=1/2 antiferromagnetic (AF) 1D quantum spin chains as essentially collective phenomenon in strongly interacting spin system and consider effects of both exchange anisotropy (EA) and staggered field (SF) [11]. The applicability of OA theory to the case of $CuGeO_3$:Fe is justified in [7] and the possible model explaining the staggered field genesis in $CuGeO_3$ doped with magnetic impurities is presented in [12]. Below we will concentrate on the OA approach to the explanation of the data in fig. 3 and alternative models will be discussed elsewhere. In the presence of the exchange anisotropy and staggered field the line width and g-factor are given by [11]

$$W(T)=W_{EA}(T) + W_{SF}(T)=A \cdot T + C \cdot h^2 \cdot T^{-2}, \quad (2a)$$

$$g(T)=g_0+g_{SF}(T)= g_0 + D \cdot h^2 \cdot T^{-3}, \quad (2b)$$

where $h$ denotes magnitude of the staggered field, and $A$, $C$ and $D$ stands for the quantities having weak temperature dependence and calculated in [11]. We wish to emphasize that in the case of low temperature growth of $W(T)$ and $g(T)$ are caused by staggered field [11] and these parameters are strongly correlated, i.e. the values $C$ and $D$ are differ by numerical coefficient [7,11]. Following the ansatz suggested in [7] we will use Eqs. (2) to define the OA function

$$f_{OA}(T)=W(T)/[\Delta g(T) \cdot T ]\equiv A^* \cdot (T^3/h^2) + C_{OA}, \quad (3)$$

where $\Delta g(T)= g(T)-g_0$ , $A^*=A/D$ and $C_{OA}$ is the universal constant in the OA theory, which does not depend on exchange integral and SF magnitude: $C_{OA}=1.99 \cdot k_B/\mu_B$ [7]. To calculate the OA function from the experimental data (fig. 3) we have used $\Delta g(T)= g(T)- g(30 K)$.

The results of our analysis are presented in fig. 4, where experimental values of $f_{OA}$ (points) are placed together with the fits using Eq. (3) (solid lines). It is visible that OA theory provides reasonable description of the experimental data in the interval $T>2$ K. However for $T<2$ K the experimental OA function starts to increase when temperature is lowered and apparently deviates from the predictions of the OA theory. This behavior may probably reflect effects of competition between interchain antiferromagnetic interactions in $CuGeO_3$ and staggered Zeeman energy [12,13]. Indeed using Mori-Kawasaki theory [14] it is possible to show that $f_{OA}$ will diverge as $(T-T_N)^{-3/4}$ [15] when temperature approach Neel point $T_N$ and hence the observed non-monotonous dependence of the $f_{OA}(T)$ may reflect transition from predominantly 1D behavior of $Cu^{2+}$ chains ($T >2$ K) to beginning of the 3D AF ordering ($T <2$ K). Further studies in the temperature range $T <0.5$ K are required to check this hypothesis.

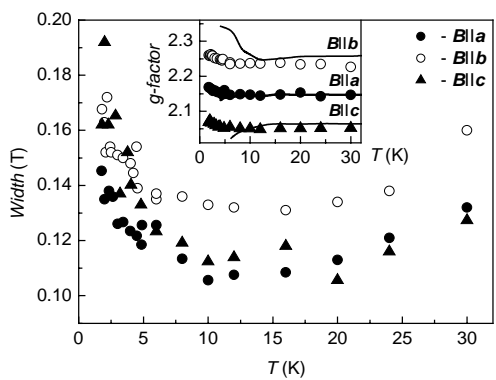

**Fig. 3**. Temperature dependences of line width and g-factor for the various orientations of magnetic field. The solid lines at inset are $g(T)$ data for the pure $CuGeO_3$ from [10].

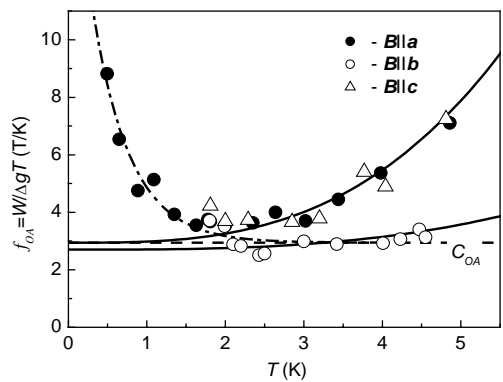

**Fig. 4**. The OA function. Points- experiment, solid lines- best fits using Eq. (3). Dashed line- universal constant in the OA theory. Dash-dot line is a guide to the eye.

Interesting that for $B \| a$ and $B \| c$ the magnitude of the coefficient at $T^3$ in Eq. (3) is almost equal and noticeably bigger than in case $B \| b$ (fig. 4). As long as in OA theory $A^*$ does not depend on magnetic field orientation, this result suggests that staggered component of the magnetic moment is located predominantly along $b$ axis. In $CuGeO_3$ structure this direction corresponds to the strongest interchain interactions.

In conclusion, we have shown that critical behavior can be observed in $CuGeO_3$:Fe in a wide range where temperature vary 40 times. The critical exponents along three crystallographic axes are very close that is in agreement with the QC model. The description of the temperature dependences of the theory. Possible arguments favoring the competition between effects of the SF and AF ordering are provided.

## ACKNOWLEDGEMENTS

Authors acknowledge support from Russian Science Support Foundation, programmes "Physics of Nanostructures" and "Strongly Correlated Electrons" of RAS and grants RFBR 04-02-16574 and INTAS 03-51-3036. This work was partly supported by Hyogo Science and Technology Association and also by Grant-in-Aid for Science Research on Priority Areas (No. 13130204) from the Ministry of Education, Culture, Sports, Science and Technology of Japan.